\documentclass{pnastwo}

\usepackage{graphicx}
\usepackage{amssymb,amsmath,amsfonts}
\usepackage{url}
\usepackage{cite}
\usepackage{array}
\usepackage{pnastwoF}
\usepackage{tabularx}
\newcolumntype{M}{>{$\vcenter\bgroup\hbox\bgroup}c<{\egroup\egroup$}}

\usepackage{ifthen}

\newcounter{MC}
\newcommand*\MC%
{%
\ifnum\theMC>0%
{MC}\fi%
\ifnum\theMC=0%
{Monte Carlo (MC)\setcounter{MC}{1}}\fi%
}

\newcounter{WHAM}
\newcommand*\WHAM%
{%
\ifnum\theWHAM>0%
{WHAM}\fi%
\ifnum\theWHAM=0%
{the weighted histogram analysis method (WHAM)\setcounter{WHAM}{1}}\fi%
}

\newcounter{SI}
\newcommand*\SI%
{%
\ifnum\theSI>0%
{SI}\fi%
\ifnum\theSI=0%
{Supplementary Information (SI)\setcounter{SI}{1}}\fi%
}

\newcounter{hcp}
\newcommand*\hcp%
{%
\ifnum\thehcp>0%
{HCP}\fi%
\ifnum\thehcp=0%
{hexagonally close-packed (HCP)\setcounter{hcp}{1}}\fi%
}

\newcounter{fcc}
\newcommand*\fcc%
{%
\ifnum\thefcc>0%
{FCC}\fi%
\ifnum\thefcc=0%
{face-centered cubic (FCC)\setcounter{fcc}{1}}\fi%
}

\newcounter{bcc}
\newcommand*\bcc%
{%
\ifnum\thebcc>0%
{BCC}\fi%
\ifnum\thebcc=0%
{body-centered cubic (BCC)\setcounter{bcc}{1}}\fi%
}

\newcounter{sc}
\newcommand*\scub%
{%
\ifnum\thesc>0%
{SC}\fi%
\ifnum\thesc=0%
{simple cubic (SC)\setcounter{sc}{1}}\fi%
}

\usepackage{verbatim}

\usepackage{youngtab}
\usepackage{tikz}
\usetikzlibrary{arrows,shapes,positioning,calc,3d,shapes.geometric}
\usepackage{makecell}

\usepackage{color}

\newcommand{\ie}{\textit{i.e.}}

\definecolor{gcb1}{cmyk}{0.35,0.07,0,0}
\definecolor{gcb2}{cmyk}{0.90,0.30,0,0}
\definecolor{gcb3}{cmyk}{0.30,0,0.45,0}
\definecolor{gcb4}{cmyk}{0.80,0,1.00,0}
\definecolor{gcb5}{cmyk}{0,0.40,0.25,0}
\definecolor{gcb6}{cmyk}{0.10,0.90,0.80,0}
\definecolor{gcb7}{cmyk}{0,0.25,0.50,0}
\definecolor{gcb8}{cmyk}{0,0.50,1.00,0}
\definecolor{gcb9}{cmyk}{0.20,0.25,0,0}
\definecolor{gcb10}{cmyk}{0.60,0.70,0,0}

\newcommand{\sharon}[1]{\textcolor{black}{#1}}

\begin{document}

\title{Shape Driven Solid--Solid Transitions in Colloids}
\author{Chrisy Xiyu Du
\affil{1}{Department of Physics,
University of Michigan, Ann Arbor, MI 48109-1040, USA}
Greg van Anders\affil{2}{Department of Chemical Engineering,
University of Michigan, Ann Arbor, MI 48109-2136, USA}
Richmond S.\ Newman\affil{2}{}
Sharon C.\ Glotzer\affil{1}{}\affil{2}{}\affil{3}{%
Department of Materials Science and Engineering,
University of Michigan, Ann Arbor, MI 48109-2136, USA}%
\affil{4}{Biointerfaces Institute,
University of Michigan, Ann Arbor, MI 48109-2800, USA}}

\contributor{Submitted to Proceedings of the National Academy of Sciences of the
United States of America}

\significancetext{%
  Despite the fundamental importance of solid--solid transitions for metallurgy,
  ceramics, earth science, reconfigurable materials, and colloidal matter, the
  details of how materials transform between two solid structures are poorly
  understood. We introduce a class of {simple} model systems in which the
  direct control of local order, via colloid shape change, induces solid--solid
  phase transitions, and characterize how the transitions happen
  thermodynamically. We find that within a single shape family there are
  solid--solid transitions that can occur with or without a thermal activation
  barrier. Our results provide new means for the study of solid--solid phase
  transitions, and have implications for designing reconfigurable materials.
}

\maketitle

\begin{article}
\begin{abstract}
  Solid-solid phase transitions are the most ubiquitous in nature, and many 
  technologies rely on them. Yet, studying them in detail is difficult because 
  of the extreme conditions (high pressure/temperature) under which many 
  such transitions occur and the high resolution equipment needed to capture 
  the intermediate states of the transformations.
  These difficulties mean that basic questions remain
  unanswered, such as whether so-called diffusionless solid--solid transitions,
  which have only only local particle rearrangement, require thermal
  activation. Here, we introduce a family of minimal model systems that exhibit
  solid--solid phase transitions that are driven by changes in the shape of
  colloidal particles.
  By using particle shape as the control variable we {entropically} reshape the
  coordination polyhedra of the particles in the system, a change {that} occurs
  indirectly in atomic solid--solid phase transitions via changes in temperature,
  pressure, or density. {We carry
  out a detailed investigation of the thermodynamics of a series of isochoric, diffusionless
  solid--solid phase transitions within a single shape family, and find {both} transitions that require
  thermal activation}, or are ``discontinuous'', and transitions that occur without
  thermal activation, or are ``continuous''. In the discontinuous case we find
  that sufficiently large
  shape changes can drive reconfiguration on time scales comparable to those for
  self-assembly, {and without an intermediate fluid phase,} 
  and in the continuous case solid--solid reconfiguration happens
  on shorter time scales than does self-assembly, providing guidance for
  developing new means of generating reconfigurable colloidal materials.
\end{abstract}

%\section{Introduction}
Despite wide-ranging implications for metallurgy \cite{portermetal}, ceramics
\cite{smith1986principles}, earth-sciences
\cite{xtalxtalearthquake,olivinespinel}, reconfigurable materials
\cite{shpphphshp,gang11}, and colloidal matter {\cite{colloidalmatter}}
fundamental questions remain about basic physical mechanisms of solid--solid
phase transitions. One major
class of solid--solid transitions is diffusionless transformations.  {Although} in
diffusionless transformations, particles undergo only local rearrangement,
the thermodynamic nature of diffusionless transitions is unclear \cite{fultz2014phase}. This gap in our understanding arises from
technical details that limit what we can learn about solid--solid transitions
from standard laboratory techniques such as X-ray diffraction or electron
microscopy \cite{nanoxtalssxfm}. The use of a broader array of experimental,
theoretical, and computational techniques could provide better understanding
of solid--solid transitions if an amenable class of 
models could be developed.\cite{quasi2d2stepssxfm} To develop
minimal models it is important to note that solid--solid transitions are
accompanied by a change in shape of the coordination polyhedra in the structure (e.g.\
\cite{Murakami855}). Coordination polyhedra reflect the bonding of atoms in a
crystal, which suggests that minimal models of solid--solid transitions could be
provided by
systems in which the shape of coordination polyhedra is directly
manipulated. {Direct} manipulation of coordination polyhedra {may be achieved
in} systems of anisotropically shaped colloids \cite{epp,entint}. Anisotropic
colloids manifest an emergent, shape-dependent entropic valence
\cite{epp,entint} that is responsible for the stabilization of a wide variety of
structures {even in the absence of \sharon{direct} interparticle forces \cite{dijkstranonconvex,trunctet,dijkstrasuperballs,escobedo,
rossi,zoopaper,geissleryang,escopoly,dijkstratcube,archimedean,epp}.} Moreover,
colloids are amenable to a wide range of observational and experimental
techniques, and colloids have been widely used to investigate
melting
\cite{premelting,hommelt}, sublimation \cite{sublimate}, crystallization
\cite{gasseretal,colloidkin,nuclkin}, and vitrification \cite{weeksglass}.
Pioneering realizations of solid--solid phase transitions in colloids
\cite{yethiraj,gang11,quasi2d2stepssxfm,quasi2dssxfm,colloidbccfcc,rossipnas}
have revealed detailed information about the {transition mechanisms}, including showing
\cite{quasi2d2stepssxfm,quasi2dssxfm} a model first-order transition that goes
through an intermediate fluid \cite{virtualmelt} and a model that showed
a variety of different transition pathways \cite{multipath}.

Here we \sharon{change} particle shape \sharon{\emph{in situ}} to directly control particle coordination in
colloidal crystals to create minimal models of solid--solid phase transitions.
Our models mimic the changes in coordination that occur indirectly in
{conventional} atomic solid--solid
transitions through changes in temperature, pressure, or density. We find
{isochoric} solid--solid transitions in ``shape space'' in a simple
two-parameter family \cite{dfamilyp} of convex colloidal polyhedra with fixed
point-group symmetry that exhibits transitions between crystals with one
[{\scub}], two [{\bcc}], and four [{\fcc}] particles in a cubic unit cell. We
study the thermodynamics of the transitions using a consistent thermodynamic
parametrization of particle shape, via the recently proposed approach of
``digital alchemy'' \cite{digitalalchemy}, combined with the rare-event sampling
technique of umbrella sampling \cite{umbrellasampling}, which is commonly
used to calculate free energy difference between two different states \cite{Kofke200541}.  We investigate
solid--solid transitions between {\bcc} and {\fcc} crystals, and between {\bcc}
and {\scub} crystals.  \sharon{We find that both} transitions are diffusionless
transformations between lattices that are continuously related by linear
mathematical transformations.  We study four cases of the
{\bcc}$\leftrightarrow${\fcc} transition and find that in all cases the
transition is thermodynamically discontinuous (i.e.\ first-order). We also study
two cases of the {\bcc}$\leftrightarrow${\scub} transition, and find, in
contrast, that in both cases the transition is thermodynamically continuous
(i.e.\ second- or higher-order). We study the dynamics of the solid--solid
transitions and find no evidence of intermediate fluid states regardless of
whether the transition is discontinuous or continuous. Our results show
  that diffusionless solid--solid phase transitions can be thermodynamically
  discontinuous or continuous, even in cases where transitions are induced by
symmetry-invariant shape change, {driven solely by entropy maximization, 
and where transitions} are between cubic crystals with
four or fewer particles in a unit cell. More generally, our approach opens up a
wide array of new systems for studying solid--solid transitions by using
particle shape to directly manipulate coordination polyhedra. Finally,
  we find systems in which solid--solid transitions occur on timescales that are
  comparable to those for self-assembly, suggesting new approaches for
rationally designing reconfigurable colloidal materials.

%%%%%%%%%%%%%Figure 1%%%%%%%%%%%%%%%%%%%
\begin{figure*}
  \begin{center}
  \includegraphics[width=0.95\textwidth]{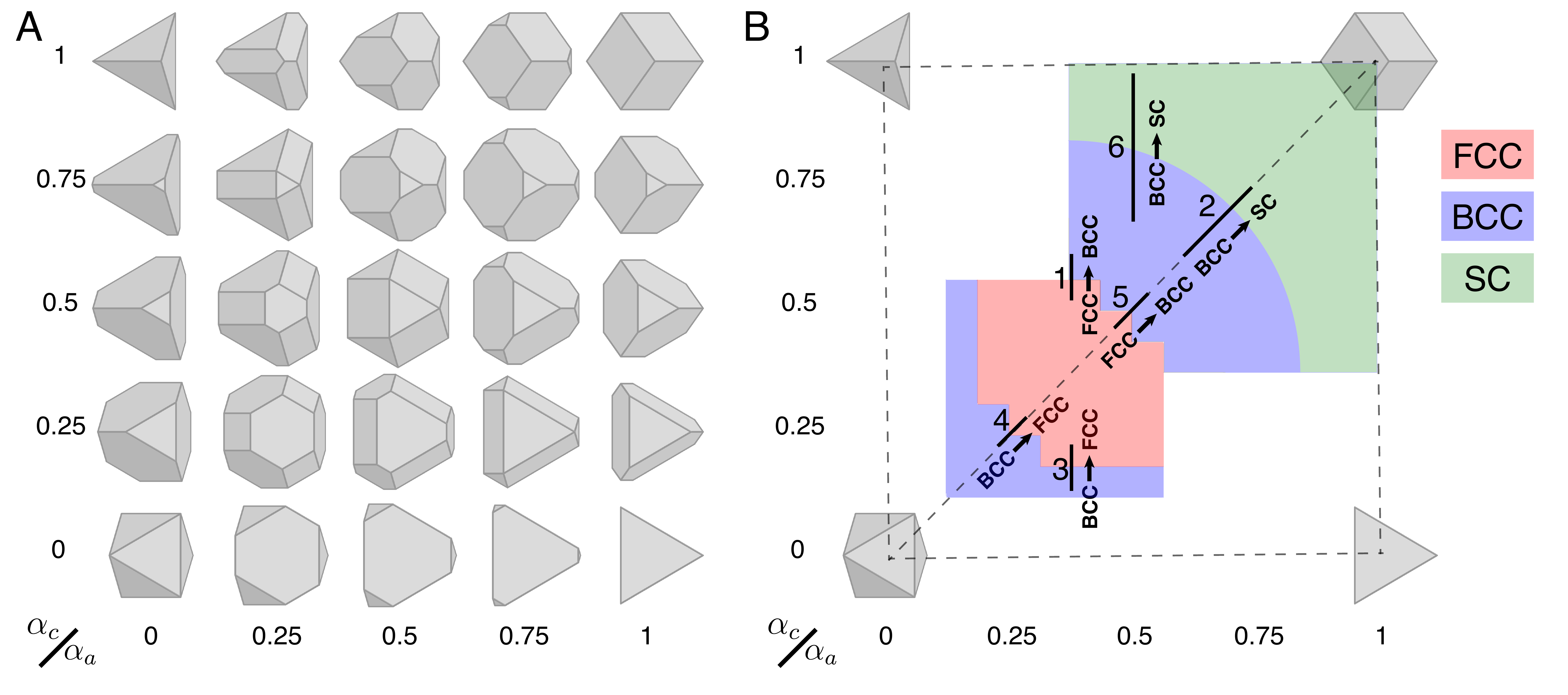}
  \end{center}
  \caption{
    (A) Spheric-triangle invariant ($\mathsf{\Delta}_{323}$) polyhedra form a
    continuous two-parameter ($\alpha_a$, $\alpha_c$) family of
    symmetric convex
    shapes that are bounded by the octahedron ($(\alpha_a,\alpha_c)=$(0,0)),
    tetrahedron ((0,1) and (1,0)) and cube (1,1). 
    (B) We show six lines indicating regions of
    {shape space} in which there is a change in the equilibrium structure at a
    packing density of $\eta=0.55$. {The lines are} annotated with the relevant
    structural transition and direction. The colors indicate the self \sharon{assembled} structures, 
    where \fcc\ is red, \bcc\ is blue and \scub\ is \sharon{green}. The self \sharon{assembled} phases indicated
    is an approximated representation from the actual phase diagram \cite{dfamilya}.  Results for \sharon{transitions} 1 and 2 are given
    in Figs.\ \ref{result1} and \ref{result2} respectively. Results for \sharon{transitions}
    3-6 are in \SI.
  \label{model}
  }
\end{figure*}

%%%%%%%%%%%%%Figure 2%%%%%%%%%%%%%%%%%%%
\begin{figure*}
  \begin{center}
  \includegraphics[width=0.95\textwidth]{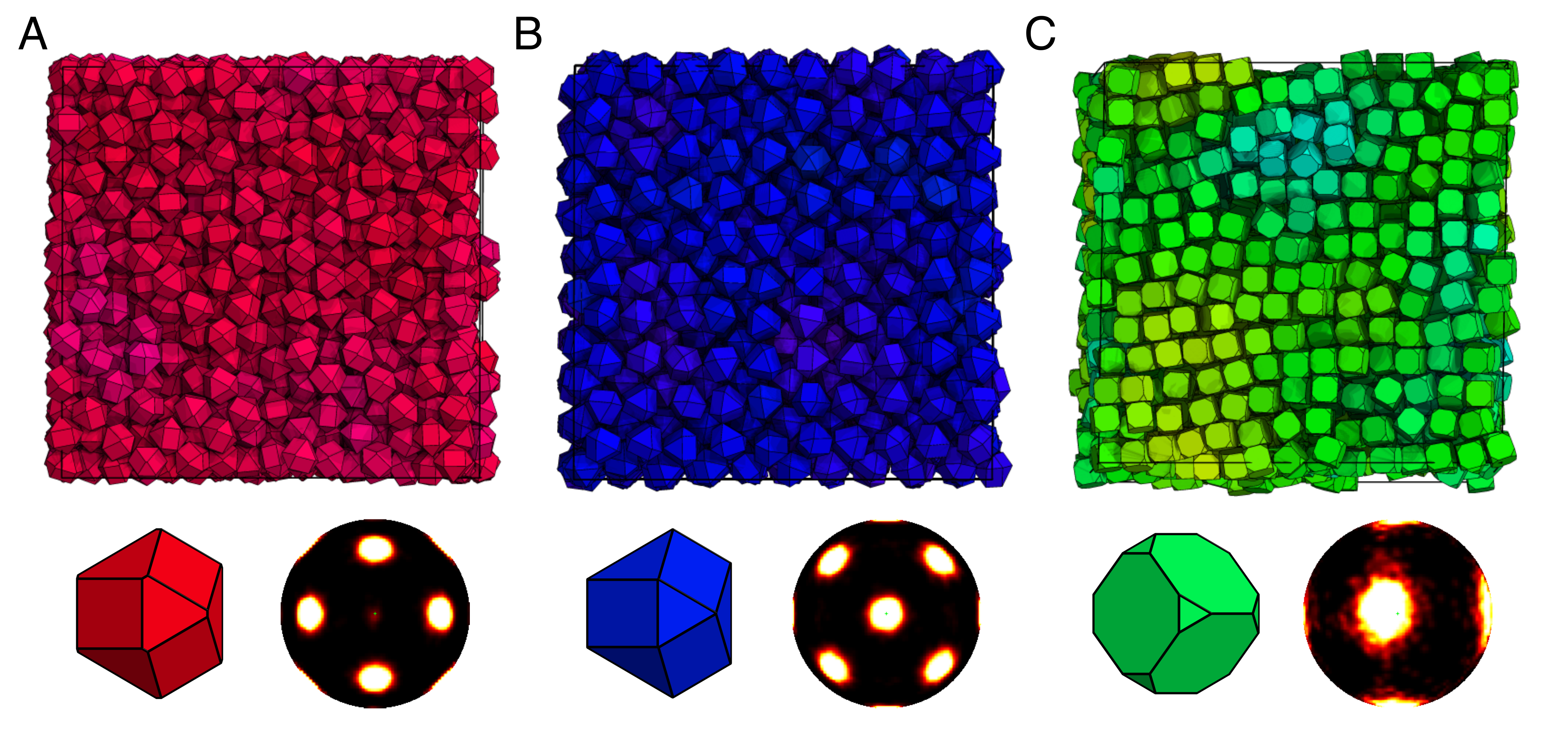}
  \end{center}
  \caption{
  Sample self assembled \sharon{colloidal crystals} formed by shapes in the
  $\mathsf{\Delta}_{323}$ triangle-invariant family of hard polyhedra, with
  images showing particle shape and bond order diagram.
  (A) An \fcc\ \sharon{crystal} self assembled from shape
  $(\alpha_a,\alpha_c)=(0.4,0.525)$.
  (B) A \bcc\ \sharon{crystal} self assembled from shape
  $(\alpha_a,\alpha_c)=(0.4,0.59)$.
  (C) A \scub\ \sharon{crystal} self assembled from shape
  $(\alpha_a,\alpha_c)=(0.76,0.76)$.
  Note the similarity of shapes in (A) and (B)\sharon{;} \sharon{even} small shape
  differences \sharon{can} affect the bulk self-assembly of hard polyhedra.  Shapes in
  (A) and (B) are both 
  on line 1 of Fig.~\ref{model} and the shape in (C) is on line 2 of Fig.~\ref{model}
  \label{sa}
  }
\end{figure*}

\section{Model and Methods}
A {family of shapes} that have the same point-group
symmetry, and {self-assemble} crystals with small 
{unit cells} (1-{\scub}, 2-{\bcc}, and 4-{\fcc}) in adjacent regions of shape space
is found in the spheric-triangle invariant 323-family ($\mathsf{\Delta}_{323}$)
of hard polyhedra \cite{dfamilyp} (see Fig.\ \ref{model} for shapes).
$\mathsf{\Delta}_{323}$ is formed by subjecting cubes to two distinct sets of
tetrahedral truncations, and includes the cube, tetrahedron, and octahedron.
These shapes, as well as intermediate shapes in $\mathsf{\Delta}_{323}$, have
been synthesized at the {colloidal scale \cite{gang11,
rossi, kayliedef, geissleryang}. Simulations} of hard colloidal polyhedra in
$\mathsf{\Delta}_{323}$ {have} shown them to have rich self-assembly behavior
(see Fig.\ \ref{sa} for three examples) with both wide and narrow regions
of thermodynamic stability for a number of different bulk structures
\cite{dijkstratcube,dfamilya}.  We denote shapes according to the conventions
$(\alpha_a,\alpha_c)$ where $0\le \alpha_{a,c}\le 1$ defines the boundaries of
shape space in this shape family. {With} these conventions $(0,0)$ is an octahedron,
$(0,1)$ and $(1,0)$ are tetrahedra \sharon{(which is self dual)}, and $(1,1)$ is a cube \sharon{(dual to the octahedron)}.
We use conventions in which all particles have
unit volume.

We investigate shape change induced solid--solid transitions in
$\mathsf{\Delta}_{323}$ in the regions indicated in Fig.\ \ref{model}, focusing
on \bcc, \fcc, and \scub\ structures. {\fcc} and {\bcc}, and {\bcc} and {\scub}
can be found in neighboring regions of $\mathsf{\Delta}_{323}$. We study
{\fcc}$\leftrightarrow${\bcc} and {\bcc}$\leftrightarrow${\scub} transitions,
and the regions of investigation indicated in Fig.\ \ref{model} comprise all
known boundaries between the phases of interest in $\mathsf{\Delta}_{323}$.

We study the thermodynamics of solid--solid transitions using both the
Ehrenfest and Landau approaches (see, e.g.\ \cite{goldenfeld}). All
{\MC ~simulations} and computations were done at fixed packing fraction $\eta=0.55$, which is
sufficiently dense to observe the spontaneous assembly of each of the target
phases \cite{dijkstratcube,dfamilya} and sufficiently dilute so as to avoid the
complicated infinite pressure behavior of this family of shapes
\cite{dijkstratcube,dfamilyp}. All simulations were performed with HOOMD-Blue
\cite{hoomdblue} using the HPMC-plugin \cite{hpmcplug}, and we use units in
which $k_\mathrm{B}T=1$.

\begin{figure}
  \vspace{-0.5cm}
  \includegraphics[width=.35\textwidth]{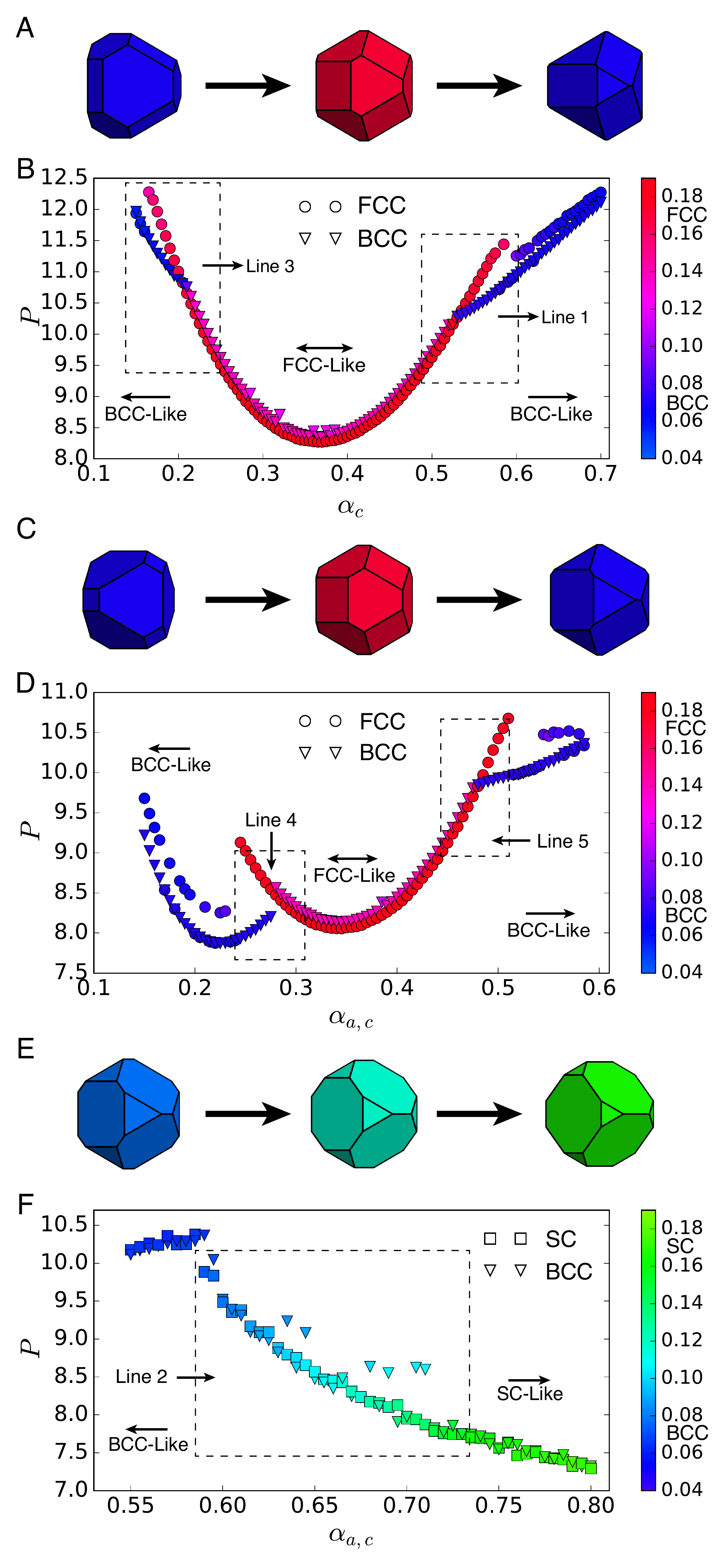}
  \vspace{0.5cm}
  \caption{
  Pressure--shape constitutive relation for $\alpha_a = 0.4$ and $\alpha_a=\alpha_c$.
  (A) Shape evolution in $\alpha_c$ at fixed $\alpha_a = 0.4$. Shapes {vary from} a \bcc\
  former (blue) to an \fcc\ former (red) and then back to a \bcc\
  former (blue).
  (B) Pressure--shape constitutive relation at fixed $\alpha_a = 0.4$.
  Circle markers indicate {\fcc} system initialization and triangle markers indicate {\bcc}
  system initialization. Marker colors indicate the value of the order parameter
  $\overline{Q}_4$
  computed in the final structure of the system after equilibration.
  Boxed regions show the \bcc$\leftrightarrow$\fcc\ boundaries corresponding to
  line 1 and line 3 in Fig.~\ref{model}. Errors are smaller than marker size.
  (C) Shape evolution for $\alpha_a = \alpha_c$. Shapes {vary from} a
  \bcc\ former (blue) to an \fcc\ former (red), then back to a \bcc\ former
  (blue).
  (D) Pressure--shape constitutive relation for $\alpha_a = \alpha_c$.
  Circle markers indicate {\fcc} system initialization; triangle 
  markers indicate {\bcc} initialization.
  (E) Shape evolution for $\alpha_a = \alpha_c$.  Shapes {vary from} a \bcc\ former (blue)
  to a \scub\ former {(green)}.
  (F) Pressure--shape constitutive relation for $\alpha_a = \alpha_c$.  Square
  markers indicate \scub\ system initialization; triangle markers indicate \bcc\ initialization.
  Marker colors indicate the value of the order parameter $\overline{Q}_4$ computed
  in the final structure of the system after equilibration.
  Boxed regions show \bcc$\leftrightarrow$\fcc\ and \bcc$\leftrightarrow$\scub\
  boundaries corresponding to lines 2, line 4 and line 5 in Fig.~\ref{model}.
  Errors are smaller than marker size.  Outliers are systems that did not equilibrate
  in 2e7 \MC\ steps.
  \label{pressure}
  }
\end{figure}

First, to estimate the location of phase boundaries, we use the notion of
generalized ``alchemical'' structure--property relationships
\cite{digitalalchemy}. Alchemical structural--property relationships are derived
from a consistent statistical mechanics treatment of particle {attributes (here, shape)} as
thermodynamic quantities. These thermodynamic quantities enter extended
$NVT\mu_a\mu_c$ ensembles that for $\mathsf{\Delta}_{323}$ have the form
\begin{equation}
  \mathcal{Z}(N,V,T,\mu_a,\mu_c) \equiv
  e^{-\beta \phi}
  = \int d\alpha_a d\alpha_c[dp][dq]e^{-\beta(H-\mu_\a N\alpha_a-\mu_c N\alpha_c)}
  \; ,
\end{equation}
where the integral is taken over shape space, as well as the ordinary 
(translational, rotational) phase
space of the particles, and where $\mu_{a,c}$ are thermodynamically conjugate to
\sharon{the shape variables} $\alpha_{a,c}$, and are referred to as ``alchemical
potentials''\cite{digitalalchemy}. The use of the term ``alchemical'' here
comes from the fact that the $\mu_{a,c}$ encode the free energy cost of
modifying the basic attributes \sharon{(here, shape)} of the particles in the system, in analogy with
the (failed) attempts of the ancient alchemists to transmute the chemical
elements. It is convenient to make a Legendre transformation of the
free energy {$\phi$} from the
$NVT\mu_a\mu_c$ ensemble to the {free energy $F=\phi+\mu_a
N\alpha_a+\mu_c N\alpha_c$} of the $NVT\alpha_a\alpha_c$ ensemble, 
from which we can extract the constitutive relation
\begin{equation}
  P(\alpha_a,\alpha_c) = -\left(\frac{\partial F}{\partial
  V}\right)_{N,T,\alpha_a,\alpha_c} \; .
\end{equation}
A thermodynamic phase transition, by the standard approach of Ehrenfest (see
e.g.\ \cite{goldenfeld})\sharon{,} is indicated if a constitutive relation, e.g.\
$P(\alpha_a,\alpha_c)$, or any of its derivatives is discontinuous. A
discontinuity of $P(\alpha_a,\alpha_c)$
signals a thermodynamic phase transition
in shape space%
{, because of the explicit shape
dependence in this relation}.
Accordingly, we searched for discontinuities by initializing systems with
different building blocks (examples in
Figs.~\ref{pressure}a,~\ref{pressure}c) in either perfect \bcc\ ($N=2000$),
\fcc\ ($N=2048$), or \scub\ ($N=2197$) structures, and \sharon{computed} 
$P(\alpha_a,\alpha_c)$ (Figs.~\ref{pressure}b,~\ref{pressure}d) after
$1.5\times10^7$ {\MC} steps to ensure systems reach equilibrium, or metastable
equilibrium, using standard techniques \cite{eppengafrenkel}.

\begin{figure}
  \vspace{-0.5cm}
  \includegraphics[width=.45\textwidth]{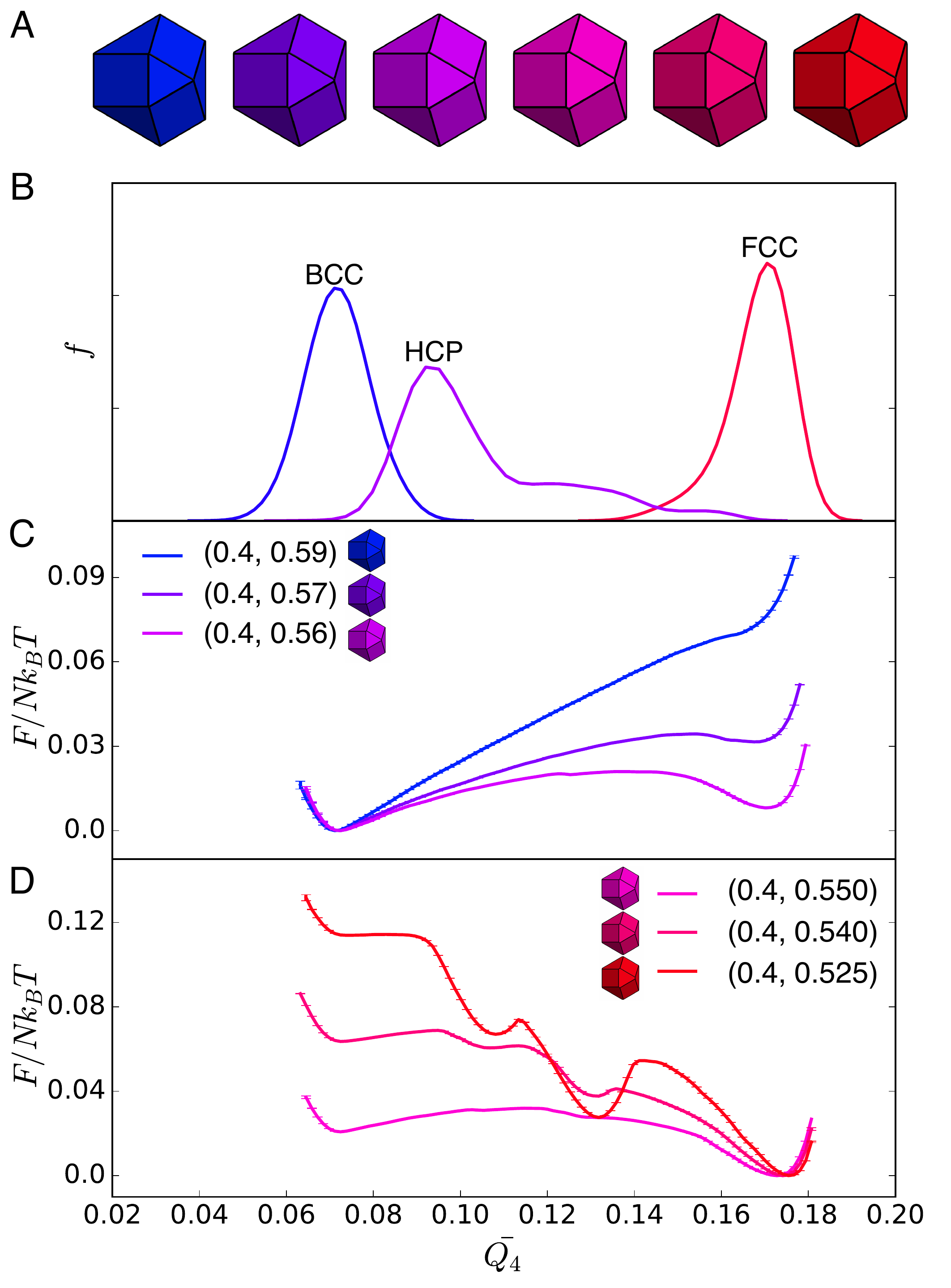}
  \vspace{0.5cm}
  \caption{
  Shape-induced structural {\fcc}$\leftrightarrow${\bcc} reconfiguration is accompanied by
  a first-order thermodynamic phase transition in spheric-triangle invariant hard
  polyhedra. 
  	(A): shapes used in umbrella sampling calculations.
   	(B): second-neighbor averaged $l=4$ spherical harmonic order
  parameter $\overline{Q}_4$ distinguishes \bcc, \fcc, and \hcp\ crystal phases in
  thermal systems of spheric-triangle invariant polyhedra. 
 	(C): above the transition ($\alpha_c > \alpha_c^*$) a metastable \fcc\ free
  energy basin develops near $\alpha_c=0.58$.  
 	(D): below the transition
  ($\alpha_c < \alpha_c^*$) the \fcc\ free energy basin becomes dominant
  ($\alpha_c=0.55$) and well above the transition ($\alpha_c=0.54$), the \bcc\ free
  energy basin becomes unstable, and a second metastable \hcp\ basin appears. 
  \label{result1}
  }
\end{figure}

\begin{figure}
  \vspace{-0.5cm}
  \includegraphics[width=.45\textwidth]{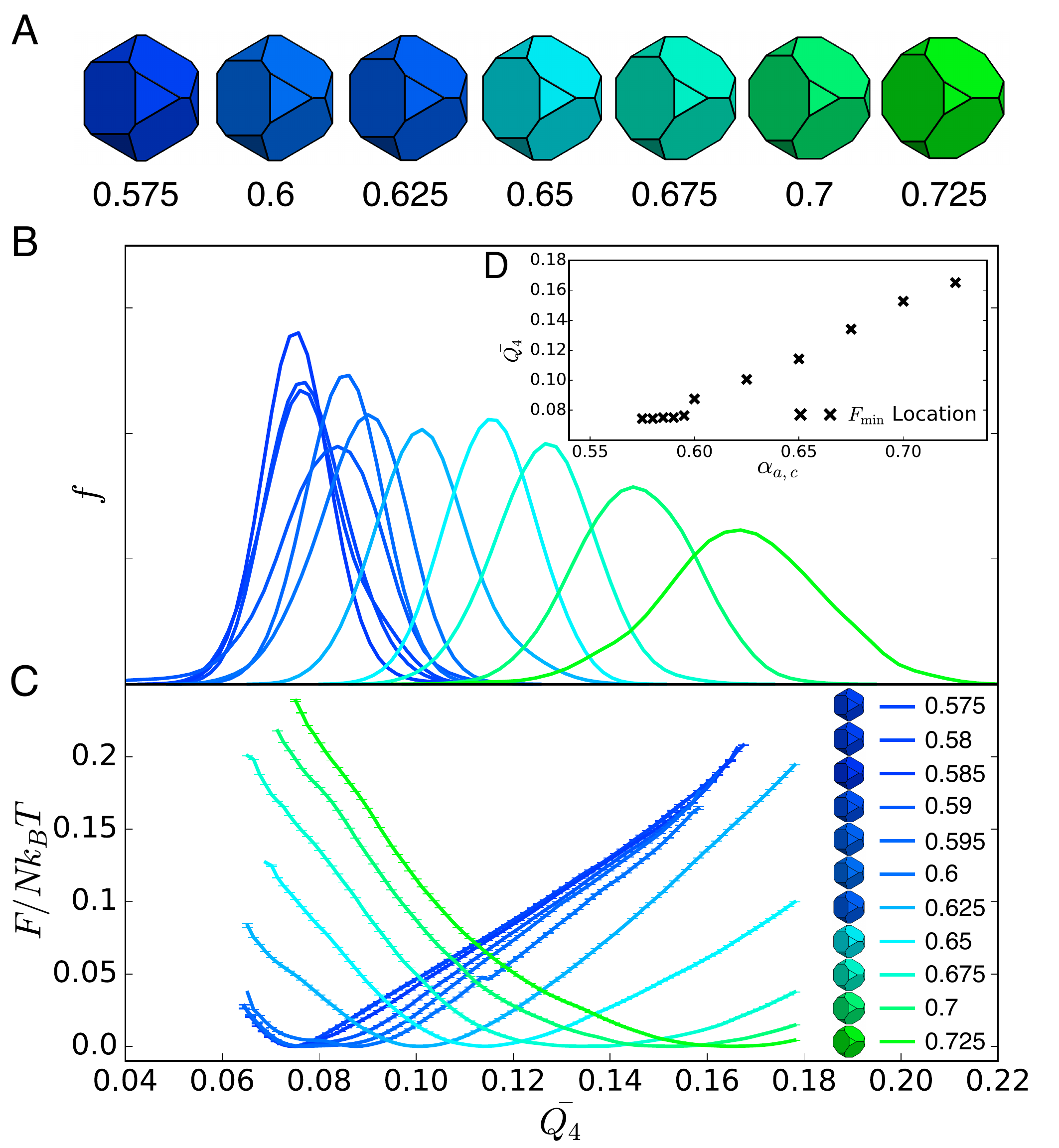}
  \vspace{0.5cm}
  \caption{
  Shape-induced structural {\bcc}$\leftrightarrow${\scub} reconfiguration occurs
  continuously
  in spheric-triangle invariant hard polyhedra. 
 	(A): sample shapes used in umbrella sampling calculation from the start to end in equal space.
		 See legend in (c) to see all shapes
  	(B): second-neighbor average $l=4$ spherical harmonic order parameter 
  $\overline{Q}_4$ shows a series of structures between {\bcc} and {\scub}.
  	(C): umbrella sampling shows a continuous
  phase transition.
  	(D): location of free energy minima extracted from umbrella sampling simulations as
		 a function of $\alpha_{a,c}$
  \label{result2}
  }
\end{figure}

Second, having located discontinuities in \sharon{$P(\alpha_a,\alpha_c)$ and its} derivatives,
we computed {the free energy as a function of order
parameter \sharon{(i.e. the Landau free energy)}} (see e.g.\
\cite{goldenfeld}), for a series of fixed particle shapes near the solid--solid
transition using umbrella sampling \cite{umbrellasampling}.  To
quantify the system \sharon{crystal} structure, we used a neighbor-averaged \cite{aveqldellago}
version of the standard spherical-harmonic bond-order parameters \cite{steinhardt}.
To achieve good order-parametric separation of our crystal phases of interest we
used the second neighbor averaged $l=4$ parameter $\overline{Q}_4$, which
distinguishes \bcc\ from \fcc, \scub, and \hcp\  phases in our systems as shown
in Figs.\ \ref{result1}c,\ref{result2}c. To confirm the
validity of $\overline{Q}_4$ as an order parameter for \sharon{monitoring the} {\fcc}$\leftrightarrow${\bcc}
\sharon{transitions,} we plot thermal averages of $\overline{Q}_4$ computed in \bcc, \hcp, and \fcc\ in
Fig.\ \ref{result1}c. Data indicate that \bcc\ crystals have a peak near
{$\overline{Q}_4=0.06$},
\hcp\ has \sharon{a} peak near {$\overline{Q}_4=0.09$} and a second smaller peak
around {$\overline{Q}_4=0.13$} due to
mixed {\hcp}--{\fcc} stacking, and \fcc\ has a peak near {$\overline{Q}_4=0.17$}.
The peaks are well separated and therefore $\overline{Q}_4$ is a good 
\sharon{distinguishing measure} of a crystal phase.  Umbrella sampling calculations used
$5\times10^4$ samples in $32$ equally-spaced windows in $\overline{Q}_4$ across each
transition with a harmonic constraint of spring constant $k=3.5\times 10^4$.
\footnote{$k$ is parametrically large because it scales like the inverse square
of the resolution of the order parameter, $\delta \overline{Q}_4$.  For our crystals
of interest, $\overline{Q}_4$ falls in the range of $0.05$ to $0.2$, so that we need
to be able to resolve order parameter intervals of $\delta
\overline{Q}_4\approx0.005$. The value $k=3.5\times 10^4$ we found to be consistent
with efficient sampling is consistent with a naive estimate $k\approx (\delta
\overline{Q}_4)^{-2}$.} We study {\fcc}$\leftrightarrow${\bcc} transitions in four
distinct regions of shape space, in each case using six polyhedra with shapes
near the solid--solid transition; all systems {contained} $N=500$
particles (Fig.~\ref{result1}a). For {\bcc}$\leftrightarrow${\scub} transitions,
we {studied} two distinct regions of shape space, in both cases using different
polyhedra in systems of $N=432$ particles (Fig.\ \ref{result2}a). In all cases,
five independent replicates \sharon{were} used to generate umbrella samples. Umbrella
samples \sharon{were} used to reconstruct free-energy curves using \WHAM\ \cite{wham}, and
errors were estimated using jackknife re-sampling \cite{jackknife}. Further
methodological details are in \SI.

%%%%%%%%%%%%%Figure 3%%%%%%%%%%%%%%%%%%%

%%%%%%%%%%%%%Figure 4%%%%%%%%%%%%%%%%%%%

%%%%%%%%%%%%%Figure 5%%%%%%%%%%%%%%%%%%%

\section{Results}
\sharon{We first present thermodynamic findings, followed by dynamics results for the
{\fcc}$\leftrightarrow${\bcc} transition.  Next, we present the same for the 
{\bcc}$\leftrightarrow${\scub} transition.}
\subsection{\sharon{{\fcc}$\leftrightarrow${\bcc} transition}}
~\ We investigated the thermodynamics of shape-change driven
{\fcc}$\leftrightarrow${\bcc} solid--solid phase transitions in four distinct
regions of shape space (indicated by lines 1, 3, 4, 5 in Fig.~\ref{model}). In
each region, at \sharon{the} {\fcc}$\leftrightarrow${\bcc} crossover\sharon{, we find that} the
$P(\alpha_a,\alpha_c)$ constitutive relation exhibits a discontinuous first
derivative (see Fig.\ \ref{pressure}b,d), indicating a phase transition that is
either first or second order in the Ehrenfest classification.\cite{goldenfeld} Further
investigation via umbrella sampling yields the Landau free energy near the
putative solid--solid transition for \sharon{the} six different shapes, depicted in Fig.\
\ref{result1}a. Note that the similarity in particle shapes makes them difficult
to distinguish by eye, but is most clearly indicated by the relative size of the
square face. Particles are colored from blue ({\bcc}) to red ({\fcc}) according
to the structures they spontaneously self-assemble. Bluer (redder) shapes
are more likely to form {\bcc} ({\fcc}). Shapes colored {purple} exhibit an almost
equal probability to form either \bcc\ or \fcc.  We computed the Landau
free energy using the order parameter $\overline{Q}_4$ defined above.  In
Fig.~\ref{result1}d,e we plot Landau free energies obtained from umbrella
sampling after averaging from five independent replica runs on both sides of the
solid--solid phase transition. Calculations at $\alpha_c>\alpha_c^*$, \ie\
above the {\fcc}$\leftrightarrow${\bcc} transition (Fig.\ \ref{result1}d), show
that sufficiently far into the \bcc\ phase there is no metastable \fcc\ free
energy basin; however, as $\alpha_c$ approaches $\alpha_c^*$, a metastable \fcc\
basin appears. At $\alpha_c<\alpha_c^*$ (Fig.\ \ref{result1}e) umbrella sampling
calculations show that \fcc\ becomes the stable free energy basin and the \bcc\
basin becomes metastable. Well below the transition the \bcc\ basin disappears,
but a metastable basin develops that corresponds to mixed \fcc\ and \hcp\
stacking. Corresponding plots that lead to the same conclusions for the other
regions of shape space are shown in \SI.  Together, our results demonstrate that
shape-change driven {\fcc}$\leftrightarrow${\bcc} solid--solid phase transitions
in $\mathsf{\Delta}_{323}$ are first-order thermodynamic phase transitions.

We investigated the dynamics of the {\fcc}$\leftrightarrow${\bcc}
solid--solid phase transition by modeling how the system responds to a sudden
change in particle shape. Several experimental techniques exist for dynamically
altering colloid shape (e.g.\ \cite{shpphphshp,lahannrecon,colloidalrecycling,
sacannashapeshift}).
Here, we model a process in which particle-shape reconfiguration occurs on a
much shorter time scale than structural relaxation by initializing
{\fcc}-forming particles in {\bcc} lattices (and vice versa) and simulating
at fixed particle shape and packing density for $1.5\times10^7$ \MC\ 
sweeps (example simulation results are shown in \SI\ movie 1). On these simulation
time scales, which are long compared to typical structural relaxation times, and
which are much longer than needed for the solid--solid transition outside
the metastable region, we did not observe structural transformation for any
systems that our umbrella sampling computation indicated as metastable.
Homogeneous nucleation is inherently a rare event, and our \MC\
simulations in metastable regions of shape space suggest that the driving force
for structural reconfiguration is not sufficient to overcome the free-energy
barrier to observe a first-order solid--solid phase transition on time scales
that are typically sufficient to observe first-order fluid--solid phase
transitions in systems of this type (e.g.\
\cite{escobedo,trunctet,zoopaper,epp}). However, we did observe a spontaneous
solid--solid transition when systems were initialized with shapes beyond the
metastable region, in which case we observed structural reconfiguration with no
discernible intermediate fluid phase. The existence of metastability in shape
space provides an additional \sharon{confirmation} that the solid--solid
transition is first order. Moreover, by measuring the order parameter evolution
in \MC\ simulation, our results indicate the transition
pathway in \MC\ simulation follows the order parameter we chose in umbrella
sampling, providing further confirmation that it appropriately parametrizes
the {\fcc}$\leftrightarrow${\bcc} solid--solid transition.

\subsection{\sharon{{\bcc}$\leftrightarrow${\scub} transition}}
\begin{figure}[t]
  \vspace{-0.5cm}
  \includegraphics[width=.45\textwidth]{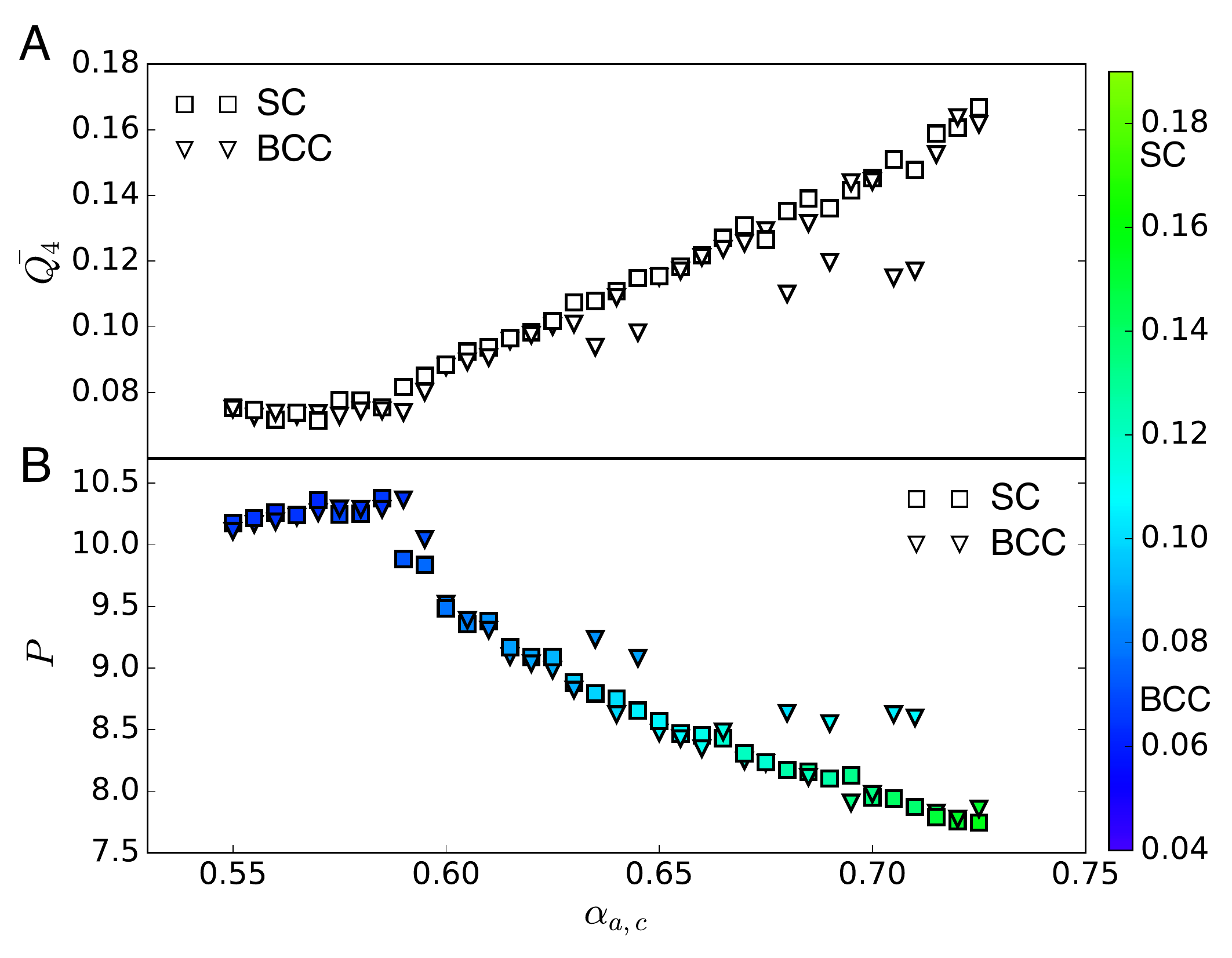}
  \vspace{0.5cm}
  \caption{
  {\bcc}$\leftrightarrow${\scub} solid--solid phase transition is a continuous
  (i.e.\ second or higher order) thermodynamic phase transition.
  (A) Order parameter $\overline{Q_4}$ versus shape suggests the derivative of the
  order parameter changes discontinuously near $\alpha_{a,c}=0.6$. (B)
  $P(\alpha_a,\alpha_c)$ also indicates a discontinuous derivative near
  $\alpha_{a,c}$ also consistent with a continuous phase transition.
  \label{result2-1}
  }
\end{figure}

~\ We {investigated} the thermodynamics of {\bcc}$\leftrightarrow${\scub}
solid--solid phase transitions in two distinct regions of shape space (lines 2,
6 in Fig.~\ref{model}). In Fig.\ \ref{pressure}d we plot the
$P(\alpha_a,\alpha_c)$ constitutive relation with
$\alpha_a=\alpha_c\equiv\alpha$ for region 2, which {shows} a discontinuity in
pressure near $\alpha\approx0.6$ consistent with a phase transition that is at
most second order in the Ehrenfest classification. A close up of these data are
presented in Fig.\ \ref{result2-1}b. Fig.\ \ref{result2-1}a shows $\overline{Q}_4$
order parameter measurements that suggest a discontinuous first derivative with
respect to $\alpha$, which is also consistent with a continuous (i.e.\ second or
higher order) thermodynamic phase transition. Corroborating evidence is provided
by computing the Landau free energy as a function of the order parameter
$\overline{Q}_4$ near the putative solid--solid transition, via umbrella sampling,
for a range of shapes indicated in Fig.\ \ref{result2}a. Particles are colored
from blue ({\bcc}) to \sharon{green} ({\scub}) according to the value of the order
parameter $\overline{Q}_4$ of the structures they \sharon{self-assemble} into. Computed
thermal averages (Fig.\ \ref{result2}b) of the order parameter $\overline{Q}_4$ in
\bcc\ and \scub\ crystals show that \bcc\ crystals have a peak near $0.08$, and
\scub\ has a peak near $0.17$; however our results also suggest the existence of
structures with intermediate $\overline{Q}_4$ for intermediate particle shapes,
where their self assembled structures are \sharon{inbetween} \bcc\ and \scub\sharon{,} as
shown in their $\overline{Q}_4$ distribution. Landau free energies computed via
umbrella sampling are plotted in Fig.\ \ref{result2}c, and show no evidence of
secondary local minima that would indicate a discontinuous (i.e.\ first order)
phase transition.  Umbrella sampling computations were performed at a higher
resolution of shape space below the putative transition ($\alpha\lesssim0.6$) to
extract the expected value of the order parameter $\overline{Q}_4$ (inset plot Fig.\
\ref{result2}d) and are consistent with the \sharon{self assembled $\overline{Q}_4$ measurements} (Fig.\ \ref{result2-1})\sharon{,}
suggesting $\overline{Q}_4$ has a discontinuous derivative at the transition.
Together, the $P(\alpha_a,\alpha_c)$ constitutive relation, the direct
evaluation of the order parameter $\overline{Q}_4$, and the umbrella sampling results
all indicate that the {\bcc}$\leftrightarrow${\scub} solid--solid phase
transition is a continuous (i.e.\ second or higher order) thermodynamic phase
transition in $\mathsf{\Delta}_{323}$. Evidence that the
{\bcc}$\leftrightarrow${\scub} solid--solid phase transition is also continuous
in shape space region 6 (see Fig.\ \ref{model}) is given in \SI.

As in the {\fcc}$\leftrightarrow${\bcc} case, we \sharon{investigated} the dynamics
of the {\bcc}$\leftrightarrow${\scub} solid--solid phase transformation by {modeling}
how the system responds to a sudden change in particle shape.  We model a
process in which particle-shape reconfiguration occurs on a much shorter
time scale than the structural relaxation by initializing {\bcc}-forming
particles in {\scub} lattices (and vice versa) and simulating at fixed particle shape and packing
density {for} $1.5\times10^7$ \MC\ sweeps (example
simulation results are shown in \SI\ movie 2). On these simulation time scales, in
all cases we observed dynamic solid--solid phase transformations, via a transition
pathway through intermediate structures that follow the order parameter we used
for umbrella sampling. Moreover, we \sharon{observe} that for $\alpha$ above the
transition any shape perturbation induces a structural change with no evidence
of metastability. We also {observe} that the dynamics of the solid--solid
transformation occurs on typical time scales of $2\times 10^6$ {\MC}
sweeps. Taken together, these results provide additional corroboration of our
observation that {\bcc}$\leftrightarrow${\scub} is a continuous thermodynamic
phase transition in $\mathsf{\Delta}_{323}$ at fixed packing fraction $\eta=0.55$,
 and the time scale under which the
solid--solid transition occurs dynamically is shorter \sharon{by nearly an order of magnitude} than in the case of
{\fcc}$\leftrightarrow${\bcc}.

\section{Discussion}
Motivated by the need for minimal models to study solid--solid
transitions \cite{quasi2d2stepssxfm}, the observation that in these transitions
coordination polyhedra change shape \cite{Murakami855}, the connection between
anisotropic colloid shape and valence \cite{epp,entint}, {the} large body of
work on entropy driven ordering in systems of colloids with anisotropic shape
\cite{dijkstranonconvex,trunctet,dijkstrasuperballs,escobedo,
rossi,zoopaper,geissleryang,escopoly,dijkstratcube,entint,epp,archimedean}, and
recently developed techniques for treating particle shape
thermodynamically \cite{digitalalchemy}, we \sharon{stuided} a
class of minimal model systems exhibiting solid--solid phase transitions driven
by changes in particle shape. We showed that particle shape change gives rise to
several distinct solid--solid transitions in a single family of shapes, and, via
{\MC} simulation and umbrella sampling techniques, we investigated
{\fcc}$\leftrightarrow${\bcc} and {\bcc}$\leftrightarrow${\scub} transitions.
Both {\fcc}$\leftrightarrow${\bcc} and {\bcc}$\leftrightarrow${\scub} are
solid--solid transitions that are related by linear transformations of the
positions of the particle centers.  \bcc\ and \fcc\ are related by elongation in
one direction \cite{bain}; \bcc\ and \scub\ are related by a \sharon{shear} transformation 
\sharon{within the unit cell \cite{ptnanocube}}.  Both
transitions are expected to be diffusionless.  Surprisingly, despite this, and
the common point-group symmetry of the particle shape, we find that the
{\fcc}$\leftrightarrow${\bcc} transition is \sharon{thermodynamically} discontinuous (i.e.\ first-order)
and {\bcc}$\leftrightarrow${\scub} is \sharon{thermodynamically} continuous (i.e.\ second or higher order).
Our results suggest several directions for further investigation.

%%%%%%%%%%%%%Figure 6%%%%%%%%%%%%%%%%%%%

The physics of {\fcc}$\leftrightarrow${\bcc} solid--solid phase transitions is of
longstanding interest in metallurgy, e.g.\ in the transition between the
so-called $\gamma$ (\fcc) and $\alpha$ (\bcc) forms of iron \cite{bain}.
We found that for several distinct regions of shape space the
{\fcc}$\leftrightarrow${\bcc} solid--solid phase transition is discontinuous.
It is straightforward to extend the methods of this work to other shape
families, and future work should determine whether our finding that the
{\fcc}$\leftrightarrow${\bcc} transition is discontinuous is the same in other
shape families. Additionally, though our focus here was on solid--solid
transitions between cubic crystals with four or fewer particles per cubic unit cell,
studies of non-cubic crystals, or crystals with more complicated unit cells, \sharon{require only}
straightforward generalizations of our approach. We expect that the approach
we have developed here will provide a powerful setting for the study of the
basic physics of solid--solid phase transitions between a wide array of
technologically relevant structures.

Constructing shape-driven solid--solid transitions furthers the aim of
developing minimal models of these transitions because it allows the direct
manipulation of coordination polyhedra. As we noted above, coordination
polyhedra also reconfigure in solid--solid transitions in metallurgy with
changes in pressure, density, or temperature. An additional complicating factor
in {those} transitions is that both enthalpy and entropy play a role, \sharon{and}
decoupling their effects is difficult \cite{fultz2014phase}. A \sharon{side}
benefit of the present approach is that in the hard particle systems we present
here, the behavior is entirely driven by entropy. \sharon{Future studies} of systems with
controllable shape and enthalpic interactions \cite{epp} could allow enthalpic
and entropic contributions to be disentangled.  An important question for further 
investigation is whether the physics of solid--solid transitions is determined by the 
structures, the particle shapes or an interplay between the two.

Another fundamental question that calls for further investigation is the
study of the kinetics of colloidal solid--solid phase transformations through
non-classical nucleation and growth. It is expected that the nucleation and
growth of solid--solid transitions will be rich because crystals break the
rotational symmetry required by classical nucleation theory, and recent
experimental evidence \cite{quasi2d2stepssxfm,quasi2dssxfmnv} shows evidence for
two-step nucleation in quasi two-dimensional systems.
Minimal colloidal models of the type constructed here provide an avenue for the
study of full, three-dimensional transformations.

\begin{figure*}
  \begin{center}
    \includegraphics[width=.95\textwidth]{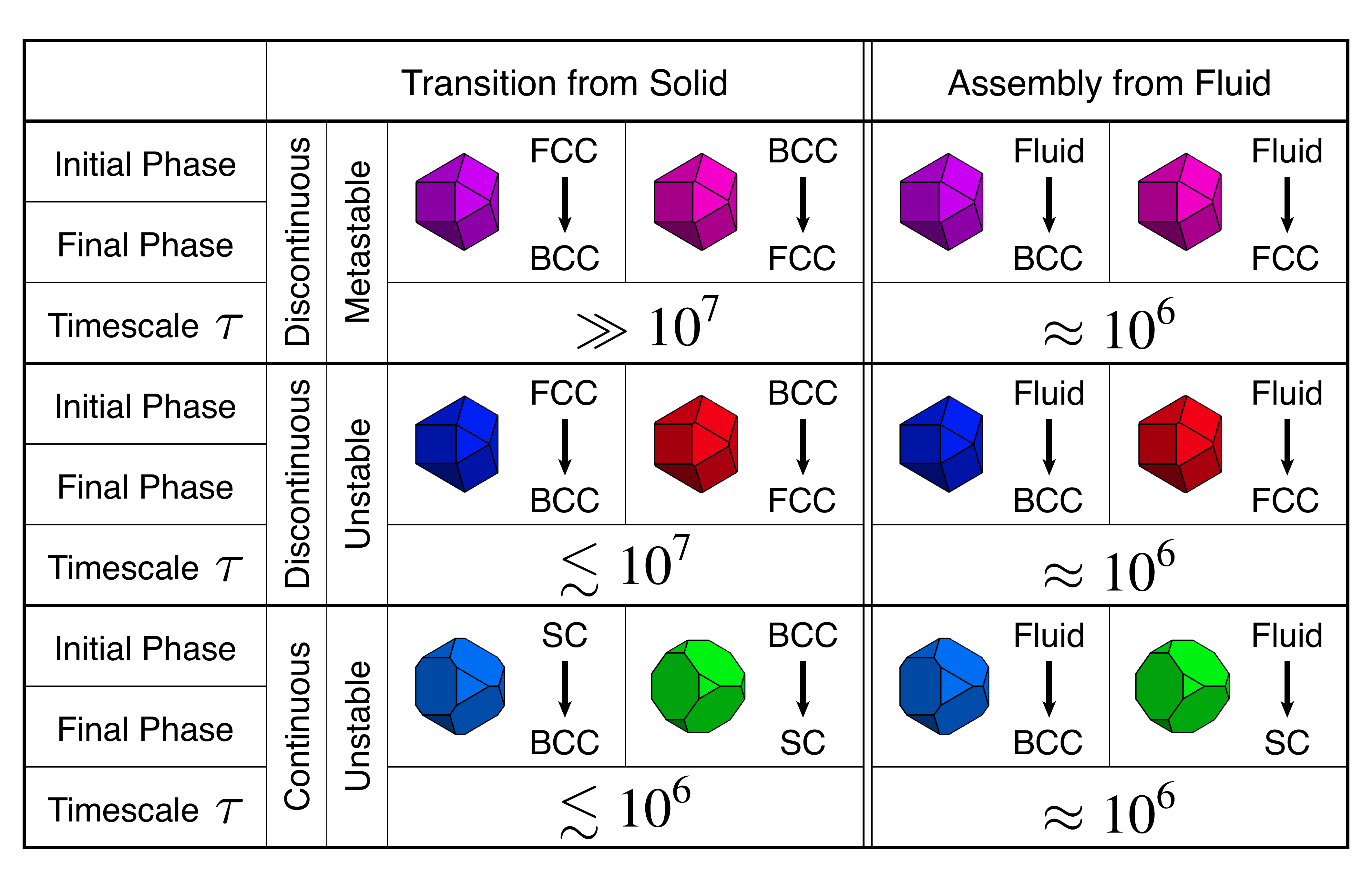}
  \end{center}
  \caption{Shape-driven solid-solid reconfiguration, and self-assembly time scales for
  \bcc\, \fcc\ and \scub\ structures.
  Thermodynamically
    discontinuous {\fcc}$\leftrightarrow${\bcc} solid--solid phase transitions
    occur dynamically in {\MC} simulations on time scales ($\tau\lesssim10^7$
    {\MC} sweeps) that are similar to self-assembly time scales
    ($\tau\approx10^6$ {\MC} sweeps) beyond the metastable region.  In the
    metastable region  solid--solid reconfiguration does not occur on time scales
    ($\tau\gg 10^7$ {\MC} sweeps) that are much longer than typical
    self-assembly times.
    Thermodynamically continuous
    {\bcc}$\leftrightarrow${\scub} solid--solid phase transitions occur
    dynamically in {\MC} simulations on time scales ($\tau\lesssim10^6$ {\MC}
    sweeps) that are comparable to or less than typical self-assembly times.
  %\caption{Shape-driven solid--solid reconfiguration, and self-assembly time
%        scales for \bcc\, \fcc\ and \scub\ structures. (a) Thermodynamically
%        discontinuous {\fcc}$\leftrightarrow${\bcc} solid--solid phase transitions
%        occur dynamically in {\MC} simulations on time scales ($\tau\lesssim10^7$
%        {\MC} sweeps) that are similar to self-assembly time scales
%        ($\tau\approx10^6$ {\MC} sweeps) beyond the metastable region.  In the
%        metastable region  solid--solid reconfiguration occurs on time scales
%        ($\tau\gg 10^7$ {\MC} sweeps) that are much longer than typical
%        self-assembly times. (b) Thermodynamically continuous
%        {\bcc}$\leftrightarrow${\scub} solid--solid phase transitions occur
%        dynamically in {\MC} simulations on time scales ($\tau\lesssim10^6$ {\MC}
%        sweeps) that are comparable to or less than typical self-assembly times.
       \label{reconfig}}
\end{figure*}

Our results can also help to guide the synthesis of reconfigurable colloidal
materials, see Fig.\ \ref{reconfig}. Experiments have demonstrated systems with
changeable building block shape, either directly
\cite{shpphphshp,lahannrecon,lahannrecon2,shapeshifting,colloidalrecycling} or
{effectively} via depletion \cite{rossipnas}. Here, we show that, for colloidal
particles that can be synthesized in the laboratory (e.g.\
\cite{rossi,geissleryang}),
changing particle shape can be used to induce transformations between 
{\fcc}$\leftrightarrow${\bcc} and {\bcc}$\leftrightarrow${\scub}. What
implications are there for the rational design of reconfigurable colloidal
materials? To answer this question it is important to understand how
structural reconfiguration compares to self-assembly in terms of typical
time scales. We obtain \sharon{``time scales''} via {\MC} simulations involving \sharon{local}
translations and rotations of individual
particles to approximate the Brownian dynamics of physical
colloids \cite{dijkstraglasspoly}. In the case
of {\fcc}$\leftrightarrow${\bcc}, for shapes near the discontinuous transition
($|\alpha-\alpha^*| \lesssim 0.05$) we did not observe spontaneous structural
reconfiguration in systems of \sharon{$N\sim 2000$} particles on time scales of
$\tau \lesssim 10^7$ {\MC} sweeps. This time scale is much longer than the
typical time it takes to observe spontaneous crystallization or melting in {\MC}
simulations of the self-assembly of $N\sim 2\times 10^3$ particles, for which
$\tau \sim 10^7$. The contrasting time scales for self-assembly versus
solid--solid reconfiguration, suggests that for small shape deformations of
$|\alpha-\alpha^*| \lesssim 0.05$, spontaneous, shape-change driven, dynamic
{\fcc}$\leftrightarrow${\bcc} reconfiguration in $\mathsf{\Delta}_{323}$ can be
achieved on shorter time scales by completely melting and then recrystallizing
the system. However, for larger shape changes $|\alpha-\alpha^*| \gtrsim 0.05$, we
observed spontaneous {\fcc}$\leftrightarrow${\bcc} reconfiguration on time scales
of $\tau \sim 10^6$ {\MC} sweeps. {This suggests that for sufficiently large shape
deformations, even though the phase transition is first-order,
direct solid--solid reconfiguration without an intermediate fluid
can occur on comparable physical time scales to self-assembly, and are therefore
a viable means of designing reconfigurable colloidal materials.} In the
{\bcc}$\leftrightarrow${\scub} case, the continuous nature of the transition
implies that there is no nucleation barrier, and, indeed, we observed structural
reconfiguration in {\MC} simulations of $N\sim2\times 10^3$ particles on typical
time scales of $\tau \sim 10^6$, which is less than what is typically observed for
self-assembly of comparably sized systems of hard, anisotropic colloids. The
relatively fast speed at which structural reconfiguration occurs in the present
case of a continuous solid--solid transition suggests that a broader search for
other systems of anisotropic colloids that exhibit continuous solid--solid phase
transitions could yield new candidate systems for developing rapidly switchable
reconfigurable colloidal materials.

\begin{acknowledgments}
  We thank K.\ Ahmed, J.\ Anderson, J.\ Dshemuchadse, O.\ Gang,
  E.\ Irrgang, and D.\ Klotsa for helpful discussions and encouragement.
  We thank D.\ Klotsa for sharing with us an early version of Ref.\ \cite{dfamilya}.
  This material is based upon work supported by, or in part by, the U.S.\ Army
  Research Office under Grant Award No.\ W911NF-10-1-0518, the DOD/ASD(R\&E)
  under Award No.\ N00244-09-1-0062, and the Department of Energy under Grant
  No.\ DE-FG02-02ER46000.  This work used the Extreme Science and Engineering 
  Discovery Environment (XSEDE), which is supported by the National Science 
  Foundation grant number ACI-1053575, XSEDE award DMR 140129. 
  Additional computational resources and services were supported by Advanced 
  Research Computing at the University of Michigan, Ann Arbor.
  X.D.\ and G.v.A.\ contributed equally to this work.
\end{acknowledgments}

\bibliographystyle{pnastwo}
\bibliography{gmaster}

\begin{thebibliography}{10}

\bibitem{portermetal}
Porter DA (2009) {\em Phase Transformations in Metals and Alloys}.
\newblock (CRC Press, Florida), 3rd edition.

\bibitem{smith1986principles}
Smith WF (1996) {\em Principles of materials science and engineering}.
\newblock (McGraw Hill, New York), 3rd edition.

\bibitem{xtalxtalearthquake}
Kirby SH, Durham WB, Stern LA (1991) Mantle phase changes and deep-earthquake
  faulting in subducting lithosphere.
\newblock {\em Science} 252(5003):216--225.

\bibitem{olivinespinel}
Burnley PC, Green HW (1989) Stress dependence of the mechanism of the
  olivine-spinel transformation.
\newblock {\em Nature} 338(6218):753--756.

\bibitem{shpphphshp}
Gang O, Zhang Y (2011) Shaping phases by phasing shapes.
\newblock {\em ACS Nano} 5(11):8459--8465.

\bibitem{gang11}
Zhang Y, Lu F, van~der Lelie D, Gang O (2011) Continuous phase transformation
  in nanocube assemblies.
\newblock {\em Phys. Rev. Lett.} 107(13):135701.

\bibitem{colloidalmatter}
Manoharan VN (2015) Colloidal matter: Packing, geometry, and entropy.
\newblock {\em Science} 349(6251):942.

\bibitem{fultz2014phase}
Fultz B (2014) {\em Phase Transitions in Materials}.
\newblock (Cambridge University Press).

\bibitem{nanoxtalssxfm}
Jacobs K, Zaziski D, Scher EC, Herhold AB, Paul~Alivisatos A (2001) Activation
  volumes for solid-solid transformations in nanocrystals.
\newblock {\em Science} 293(5536):1803--1806.

\bibitem{quasi2d2stepssxfm}
Peng Y, Wang F, Wang Z, Alsayed AM, Zhang Z, Yodh AG, Han Y (2015) Two-step
  nucleation mechanism in solid--solid phase transitions.
\newblock {\em Nat Mater} 14(1):101--108.

\bibitem{Murakami855}
Murakami M, Hirose K, Kawamura K, Sata N, Ohishi Y (2004) Post-perovskite phase
  transition in mgsio3.
\newblock {\em Science} 304(5672):855--858.

\bibitem{epp}
van Anders G, Ahmed NK, Smith R, Engel M, Glotzer SC (2014) Entropically patchy
  particles: Engineering valence through shape entropy.
\newblock {\em ACS Nano} 8:931--940.

\bibitem{entint}
van Anders G, Klotsa D, Ahmed NK, Engel M, Glotzer SC (2014) Understanding
  shape entropy through local dense packing.
\newblock {\em Proc. Natl. Acad. Sci. U.S.A.} 111(45):E4812--E4821.

\bibitem{dijkstranonconvex}
de~Graaf J, van Roij R, Dijkstra M (2011) Dense regular packings of irregular
  nonconvex particles.
\newblock {\em Phys. Rev. Lett.} 107(15):155501.

\bibitem{trunctet}
Damasceno PF, Engel M, Glotzer SC (2012) {Crystalline Assemblies and Densest
  Packings of a Family of Truncated Tetrahedra and the Role of Directional
  Entropic Forces}.
\newblock {\em ACS Nano} 6(1):609--614.

\bibitem{dijkstrasuperballs}
Ni R, Gantapara AP, de~Graaf J, van Roij R, Dijkstra M (2012) Phase diagram of
  colloidal hard superballs: from cubes via spheres to octahedra.
\newblock {\em Soft Matter} 8(34):8826--8834.

\bibitem{escobedo}
Agarwal U, Escobedo FA (2011) Mesophase behaviour of polyhedral particles.
\newblock {\em Nat. Mater.} 10(3):230--235.

\bibitem{rossi}
Rossi L, Sacanna S, Irvine WTM, Chaikin PM, Pine DJ, Philipse AP (2011) Cubic
  crystals from cubic colloids.
\newblock {\em Soft Matter} 7(9):4139--4142.

\bibitem{zoopaper}
Damasceno PF, Engel M, Glotzer SC (2012) {Predictive Self-Assembly of Polyhedra
  into Complex Structures}.
\newblock {\em Science} 337(6093):453--457.

\bibitem{geissleryang}
Henzie J, Gr{\"u}nwald M, Widmer-Cooper A, Geissler PL, Yang P (2012)
  Self-assembly of uniform polyhedral silver nanocrystals into densest packings
  and exotic superlattices.
\newblock {\em Nat. Mater.} 11:131--137.

\bibitem{escopoly}
Agarwal U, Escobedo FA (2012) Effect of quenched size polydispersity on the
  ordering transitions of hard polyhedral particles.
\newblock {\em J. Chem. Phys.} 137(2):024905.

\bibitem{dijkstratcube}
Gantapara AP, de~Graaf J, van Roij R, Dijkstra M (2013) Phase diagram and
  structural diversity of a family of truncated cubes: Degenerate close-packed
  structures and vacancy-rich states.
\newblock {\em Phys. Rev. Lett.} 111(1):015501.

\bibitem{archimedean}
Millan JA, Ortiz D, van Anders G, Glotzer SC (2014) Self-assembly of
  archimedean tilings with enthalpically and entropically patchy polygons.
\newblock {\em ACS Nano} 8(3):2918--2928.

\bibitem{premelting}
Alsayed AM, Islam MF, Zhang J, Collings PJ, Yodh AG (2005) Premelting at
  defects within bulk colloidal crystals.
\newblock {\em Science} 309(5738):1207--1210.

\bibitem{hommelt}
Wang Z, Wang F, Peng Y, Zheng Z, Han Y (2012) Imaging the homogeneous
  nucleation during the melting of superheated colloidal crystals.
\newblock {\em Science} 338(6103):87--90.

\bibitem{sublimate}
Savage JR, Blair DW, Levine AJ, Guyer RA, Dinsmore AD (2006) Imaging the
  sublimation dynamics of colloidal crystallites.
\newblock {\em Science} 314(5800):795--798.

\bibitem{gasseretal}
Gasser U, Weeks ER, Schofield A, Pusey PN, Weitz DA (2001) Real-space imaging
  of nucleation and growth in colloidal crystallization.
\newblock {\em Science} 292(5515):258--262.

\bibitem{colloidkin}
Anderson VJ, Lekkerkerker HNW (2002) Insights into phase transition kinetics
  from colloid science.
\newblock {\em Nature} 416(6883):811--815.

\bibitem{nuclkin}
Tan P, Xu N, Xu L (2014) Visualizing kinetic pathways of homogeneous nucleation
  in colloidal crystallization.
\newblock {\em Nat Phys} 10(1):73--79.

\bibitem{weeksglass}
Weeks ER, Crocker JC, Levitt AC, Schofield A, Weitz DA (2000) Three-dimensional
  direct imaging of structural relaxation near the colloidal glass transition.
\newblock {\em Science} 287(5453):627--631.

\bibitem{yethiraj}
Yethiraj A, Wouterse A, Groh B, van Blaaderen A (2004) Nature of an
  electric-field-induced colloidal martensitic transition.
\newblock {\em Phys. Rev. Lett.} 92(5):058301.

\bibitem{quasi2dssxfm}
Qi W, Peng Y, Han Y, Bowles RK, Dijkstra M (2015) Nonclassical nucleation in a
  solid-solid transition of confined hard spheres.
\newblock {\em Phys. Rev. Lett.} 115(18):185701.

\bibitem{colloidbccfcc}
Mohanty PS, Bagheri P, N{\"o}jd S, Yethiraj A, Schurtenberger P (2015) Multiple
  path-dependent routes for phase-transition kinetics in thermoresponsive and
  field-responsive ultrasoft colloids.
\newblock {\em Phys. Rev. X} 5(1):011030.

\bibitem{rossipnas}
Rossi L, Soni V, Ashton DJ, Pine DJ, Philipse AP, Chaikin PM, Dijkstra M,
  Sacanna S, Irvine WTM (2015) Shape-sensitive crystallization in colloidal
  superball fluids.
\newblock {\em Proc. Natl. Acad. Sci. U.S.A.} 112(17):5286--5290.

\bibitem{virtualmelt}
Levitas VI, Henson BF, Smilowitz LB, Asay BW (2004) Solid-solid phase
  transformation via virtual melting significantly below the melting
  temperature.
\newblock {\em Phys. Rev. Lett.} 92(23):235702.

\bibitem{multipath}
Mohanty PS, Bagheri P, N{\"o}jd S, Yethiraj A, Schurtenberger P (2015) Multiple
  path-dependent routes for phase-transition kinetics in thermoresponsive and
  field-responsive ultrasoft colloids.
\newblock {\em Phys. Rev. X} 5(1):011030.

\bibitem{dfamilyp}
Chen ER, Klotsa D, Engel M, Damasceno PF, Glotzer SC (2014) Complexity in
  surfaces of densest packings for families of polyhedra.
\newblock {\em Phys. Rev. X} 4(1):011024.

\bibitem{digitalalchemy}
van Anders G, Klotsa D, Karas AS, Dodd PM, Glotzer SC (2015) {Digital Alchemy
  for Materials Design: Colloids and Beyond}.
\newblock {\em ACS Nano} 9:9542--9553.

\bibitem{umbrellasampling}
Torrie G, Valleau J (1977) Nonphysical sampling distributions in monte carlo
  free-energy estimation: Umbrella sampling.
\newblock {\em J. Comp. Phys.} 23(2):187 -- 199.

\bibitem{Kofke200541}
Kofke DA (2005) Free energy methods in molecular simulation.
\newblock {\em Fluid Phase Equilibria} 228–229:41 -- 48.
\newblock \{PPEPPD\} 2004 Proceedings.

\bibitem{kayliedef}
Young KL, Personick ML, Engel M, Damasceno PF, Barnaby SN, Bleher R, Li T,
  Glotzer SC, Lee B, Mirkin CA (2013) A directional entropic force approach to
  assemble anisotropic nanoparticles into superlattices.
\newblock {\em Angew. Chem., Int. Ed.} 52:13980--13984.

\bibitem{dfamilya}
Klotsa D, Chen ER, Engel M, Damasceno PF, Glotzer SC (2016) To appear.
\newblock {\em In preparation}.

\bibitem{goldenfeld}
Goldenfeld N (1992) {\em Lectures on phase transitions and the renormalization
  group}.
\newblock (Addison-Wesley, Reading MA).

\bibitem{hoomdblue}
Anderson JA, Glotzer SC (2013) {The Development and Expansion of HOOMD-Blue
  Through Six Years of GPU Proliferation}.
\newblock {\url{http://codeblue.umich.edu/hoomd-blue}}.

\bibitem{hpmcplug}
Anderson JA, Irrgang ME, Glotzer SC (2016) Scalable metropolis monte carlo for
  simulation of hard shapes.
\newblock {\em Comp. Phys. Commun.} 204:21 -- 30.

\bibitem{eppengafrenkel}
Eppenga R, Frenkel D (1984) Monte carlo study of the isotropic and nematic
  phases of infinitely thin hard platelets.
\newblock {\em Molecular Physics} 52:1303--1334.

\bibitem{aveqldellago}
Lechner W, Dellago C (2008) Accurate determination of crystal structures based
  on averaged local bond order parameters.
\newblock {\em J. Chem. Phys.} 129(11):114707.

\bibitem{steinhardt}
Steinhardt PJ, Nelson DR, Ronchetti M (1983) Bond-orientational order in
  liquids and glasses.
\newblock {\em Phys. Rev. B} 28(2):784--805.

\bibitem{wham}
Kumar S, Rosenberg JM, Bouzida D, Swendsen RH, Kollman PA (1992) The weighted
  histogram analysis method for free-energy calculations on biomolecules. i.
  the method.
\newblock {\em J. Comput. Chem.} 13(8):1011--1021.

\bibitem{jackknife}
Efron B (1981) Nonparametric estimates of standard error: The jackknife, the
  bootstrap and other methods.
\newblock {\em Biometrika} 68(3):589--599.

\bibitem{lahannrecon}
Lee KJ, Yoon J, Rahmani S, Hwang S, Bhaskar S, Mitragotri S, Lahann J (2012)
  Spontaneous shape reconfigurations in multicompartmental microcylinders.
\newblock {\em Proc. Natl. Acad. Sci. U.S.A.} 109(40):16057--16062.

\bibitem{colloidalrecycling}
Meester V, Verweij RW, van~der Wel C, Kraft DJ (2016) Colloidal recycling:
  Reconfiguration of random aggregates into patchy particles.
\newblock {\em ACS Nano} 10(4):4322--4329.

\bibitem{sacannashapeshift}
Youssef M, Hueckel T, Yi GR, Sacanna S (2016) Shape-shifting colloids via
  stimulated dewetting.
\newblock {\em Nat Commun} 7:12216.

\bibitem{bain}
Bain EC, Dunkirk NY (1924) {The Nature of Martensite}.
\newblock {\em Trans. AIME} 70(1):25--47.

\bibitem{ptnanocube}
Li R, Bian K, Wang Y, Xu H, Hollingsworth JA, Hanrath T, Fang J, Wang Z (2015)
  An obtuse rhombohedral superlattice assembled by pt nanocubes.
\newblock {\em Nano Letters} 15(9):6254--6260.

\bibitem{quasi2dssxfmnv}
Sanz E, Valeriani C (2015) Crystal-crystal transitions: Mediated by a liquid.
\newblock {\em Nat Mater} 14(1):15--16.

\bibitem{lahannrecon2}
Saha S, Copic D, Bhaskar S, Clay N, Donini A, Hart AJ, Lahann J (2012)
  Chemically controlled bending of compositionally anisotropic microcylinders.
\newblock {\em Angew. Chem., Int. Ed.} 51(3):660--665.

\bibitem{shapeshifting}
Zhou J, Turner SA, Brosnan SM, Li Q, Carrillo JMY, Nykypanchuk D, Gang O, Ashby
  VS, Dobrynin AV, Sheiko SS (2014) Shapeshifting: Reversible shape memory in
  semicrystalline elastomers.
\newblock {\em Macromolecules} 47(5):1768--1776.

\bibitem{dijkstraglasspoly}
Tasios N, Gantapara AP, Dijkstra M (2014) Glassy dynamics of convex polyhedra.
\newblock {\em J. Chem. Phys.} 141(22):224502.

\end{thebibliography}
\end{article}
\end{document}